# Computational Fluid Dynamic Approach for Biological System Modeling

Weidong Huang[1]; Chundu Wu[2]; Bingjia Xiao[3] and Weidong Xia[4]
[1]School of Earth and Space Science, and [4] School of Engineering Science, University of Science and Technology of China, Hefei, China 230026
[2]School of life and environ. engineering, Jiangsu University, Zhenjiang, Jiangsu, China, 212013
[3] Institute of Plasma Physics, Chinese Academy of Sciences, Hefei, China 230031
[1] correspond author, email: huangwd@ustc.edu.cn

**Abstracts:**

**Various biological system models have been proposed in systems biology, which are based on the complex biological reactions kinetic of various components. These models are not practical because we lack of kinetic information. In this paper, it is found that the enzymatic reaction and multi-order reaction rate is often controlled by the transport of the reactants in biological systems. A Computational Fluid Dynamic (CFD) approach, which is based on transport of the components and kinetics of biological reactions, is introduced for biological system modeling. We apply this approach to a biological wastewater treatment system for the study of metabolism of organic carbon substrates and the population of microbial. The results show that CFD model coupled with reaction kinetics is more accurate and more feasible than kinetic models for biological system modeling.**

## Introduction

Systems biology integrates knowledge from diverse biological components and data into models of the system as a whole to investigate the behavior and relationships of all elements in a particular biological system[1]. To establish a modeling approach is one of the key tasks for systems biology[2-3]. Present models of biological system are most commonly based on the framework of deterministic chemical kinetics. This approach applied to study biological system has already provided valuable results, for example in studies of the networks controlling bacterial chemotaxis[6-7], developmental patterning in Drosophila[8-9] and infection of E. coli by lambda phage[10].

The major difficulty in applying deterministic chemical kinetics to modeling biological system is that we do not have enough information about the chemical kintics[4,5]. Further more, the slow transport of reactants may influence the reaction rate remarkably, for the biological reaction, especially the enzymatic reaction, is very fast in comparison with the transport of reactants, so the transport behavior of the components in a biological system should be considered in the model[11].

In this paper, a computational fluid dynamic approach is introduced for modeling of biological system. It is based on the transport of the components in biological systems with the inclusion of reaction kinetics. The population of the microbial and metabolism of organic carbon substrates and nitrogen in biological system for wastewater treatment are studied with the CFD approach. Firstly the approach is described with basic biological reactions, then, a benchmark biological wastewater treatment system[12] is simulated.

Computational fluid dynamics is a body of knowledge and techniques used to solve mathematical models of fluid dynamics on digital computers[19]. This formulates and solves the fundamental mass, momentum, and energy conservation equations in space. For many processes it is, in principle, possible to incorporate a diverse range of phenomena within these basic equations and/or to complement them with additional conservations such as biological reactions. Examples of this approach applied in biological system have recently been provided for flow and mass transfer study by many researchers[20-22].

We focus on the aerobic biological system for removal of organic carbon substances and nitrogen in wastewater, which is the main objective of the biological wastewater treatment system[23]. This kind of biological system is a man-controlled ecological system, first developed at early 20th century and remained one of the most important wastewater treatment systems[24]. The ecological system, in a constant stirred reactor with a cylinder dimension of radii and height of 4m, using high speed surface aerator in top of the center[25], is modeled with the method to study the population of the microbial and metabolism of the substrates and nitrogen.

## Method

The mass transfer of the components in the system with biological reaction is controlled by convection and diffusion in the water, and the formula control the mass transfer of a component is as following:

$$\rho(\frac{\partial C}{\partial t} + u_x \frac{\partial C}{\partial x} + u_y \frac{\partial C}{\partial y} + u_z \frac{\partial C}{\partial z}) = D(\frac{\partial^2 C}{\partial x^2} + \frac{\partial^2 C}{\partial y^2} + \frac{\partial^2 C}{\partial z^2}) + S + \sum Ri \qquad 1$$

C represents the concentration of the component, D is the diffusion coefficient and S the external supplied source term. ∑Ri represents the source term produced by all biological reactions related to the component. The velocity u can be solved by a CFD software. In the CFD modeling, A steady-state simulation is performed, because steady-state simulation greatly reduces the complexity and computer time, and demonstrates the main features of the system. It is true that the model kinetic parameters are temperature dependent, but temperature variation is so little that it can be ignored within the system, and thus, the energy equation is not necessary to be solved.

In the present kinetic model, a homogenous distribution of each component has been assumed, so that the convection and diffusion terms in Formula 1 are ignored. In actual biological system such as the biological wastewater treatment system, it has been demonstrated that distribution of oxygen is not homogenous and the transport of the components cannot be ignored in our previous work[25].

In the removal of readily biodegradable substrate, such as glucose, two biological reactions are related to the removal of substrate and growth of heterotrophs in the biological wastewater treatment system[26]. They are coupled to CFD model as following:

$C_5H_7O_2N$ +(1/6) $C_6H_{12}O_6$+0.29$O_2$+0.142$NH_3$==1.142$C_5H_7O_2N$ + 0.29 $CO_2$ + 0.716 $H_2O$ (1)

$C_5H_7O_2N$ + 2.76$H_2O$ == 0.08 $C_5H_7O_2N$ (d) + (4.6/6)$C_6H_{12}O_6$ + 0.92$NH_3$ (2)

Reaction (1) describes the growth of the heterotrophs ($C_5H_7O_2N$) and removal of the soluble organic carbon substrates ($C_6H_{12}O_6$), Reaction (2) describes the death of the heterotrophs and decomposition as organic carbon substrate and some inert residues ($C_5H_7O_2N$ (d)).

A Monad type kinetic model[27] is used for aerobic growth of heterogrophs. The main kinetics formula in the system should be that:
The heterotrophs growth rate R(X) is proportional to the heterotrophs concentration $X_{B,H}$:

$$R(X)=\mu_H X_{B,H} \quad 2$$

$$\mu_H = \mu_H^0 \frac{C_S}{C_S + K_S} \bullet \frac{C_{O_2}}{C_{O_2} + K_{O_2}} \quad 3$$

Removal rate R(S) of soluble organic carbon substrate

$$R(S)= -\frac{\mu_H}{Y_H} X_{B,H} \quad 4$$

Oxygen consumption:

$$R(O2)= -\frac{1-Y_H}{Y_H} \mu_H X_{B,H} \quad 5$$

Where $C_s$ represents the substrate concentration, $C_{O2}$ represents the dissolved oxygen concentration. The death of the heterotroph is modeled according to the death–regeneration concept[28]. The death rate R(XD) for heterotrophs in system is

$$R(XD)=-b_H \bullet X_{B,H} \quad 6$$

The producing rate R(SD) for COD due to the death of the heterotrophs:

$$R(SD)= b_H \bullet X_{B,H} \bullet (1-f) \quad 7$$

The concentration of NH3, having a little influence on the reaction rates at high concentration in actual reactor, is ignored in simulation. The generation and removal of the three components including dissolved oxygen, substrate and heterotroph are controlled by the Eqs.2-7 which has been implemented in the transport calculation. The main kinetics parameters used in the model for comparing with kinetic model are listed in the table 1.

Table1 Kinetic parameters used in the demonstration of the approach

| *Parameter* | *symbol* | *Unit* | *Data* |
|---|---|---|---|
| Heterotrophic max. specific growth rate | $\mu^0_H$ | $Day^{-1}$ | *12* |
| Heterotrophic decay rate | $b_H$ | $Day^{-1}$ | *0.40* |
| Half-saturation coefficient(hsc) of substrate for heterotrophs | $K_S$ | $gCODm^{-3}$ | *10* |
| Oxygen hsc for heterotrophs | $K_{O2}$ | $gO2m^{-3}$ | *0.2* |
| Heterotrophic yield | $Y_H$ | | *0.71* |
| Fraction of biomass yielding particulate products | *f* | | *0.08* |

Fluent6 version code[29] is employed for the transport calculation. This code uses the Finite Volume method[19] for the discretization of the Navier-Stokes equations. Standard κ-ε model[30] is used to solve the turbulent flow, which has proved rather successful for flow and mass transfer modeling in our case[25]. the reactor is cylindrically symmetric, the computational domain is simplified as two dimension. The initial grid generated for CFD calculation is about 100*100, and adapted automatically by the Fluent6 code until a convergence criterion, i.e. 5%, is achieved. For each size of the grid, the convergence criterion for each variable is set to be below $10^{-6}$. The model is robust, accurate and cost effective in terms of computational time in a similar case[31]. Detailed boundary condition can be found in previous study[25]. All designs are implemented on a personal computer with Athonon XP CPU. Because the mass transfer in our system have little influence on the flow field, the flow field is firstly solved without consideration of mass transfer, and then coupled calculation for fluid flow and mass transfer is performed, so that the calculation is fast and easy to converge.

In the perfectly mixed system assumption of kinetic model, the components in the outflow is the same as those in the system, so

$$V*dX/dt = Q*X^0_{B,H} - Q* X_{B,H} + R(X) + R(XD) \quad 8$$

$$V*dCs/dt = Q*C^0_s - Q*C_s + R(S) + R(SD) \quad 9$$

$X^0_{B,H}$ and $C^0_S$ represent the inflow concentration of the microbial and the substrate. The following formula for $X_{B,H}$ and Cs in the system can be deduced under steady state:

$$\tau = \frac{C_s^0 - C_s}{X^0_{B,H}(\mu_H / Y - f) + (C_s^0 - C_s)(\mu_H - b)} = \frac{V}{Q} \quad 10$$

$$X_{B,H} = \frac{1}{1 + (b - \mu_H)\tau} \quad 11$$

**Results**

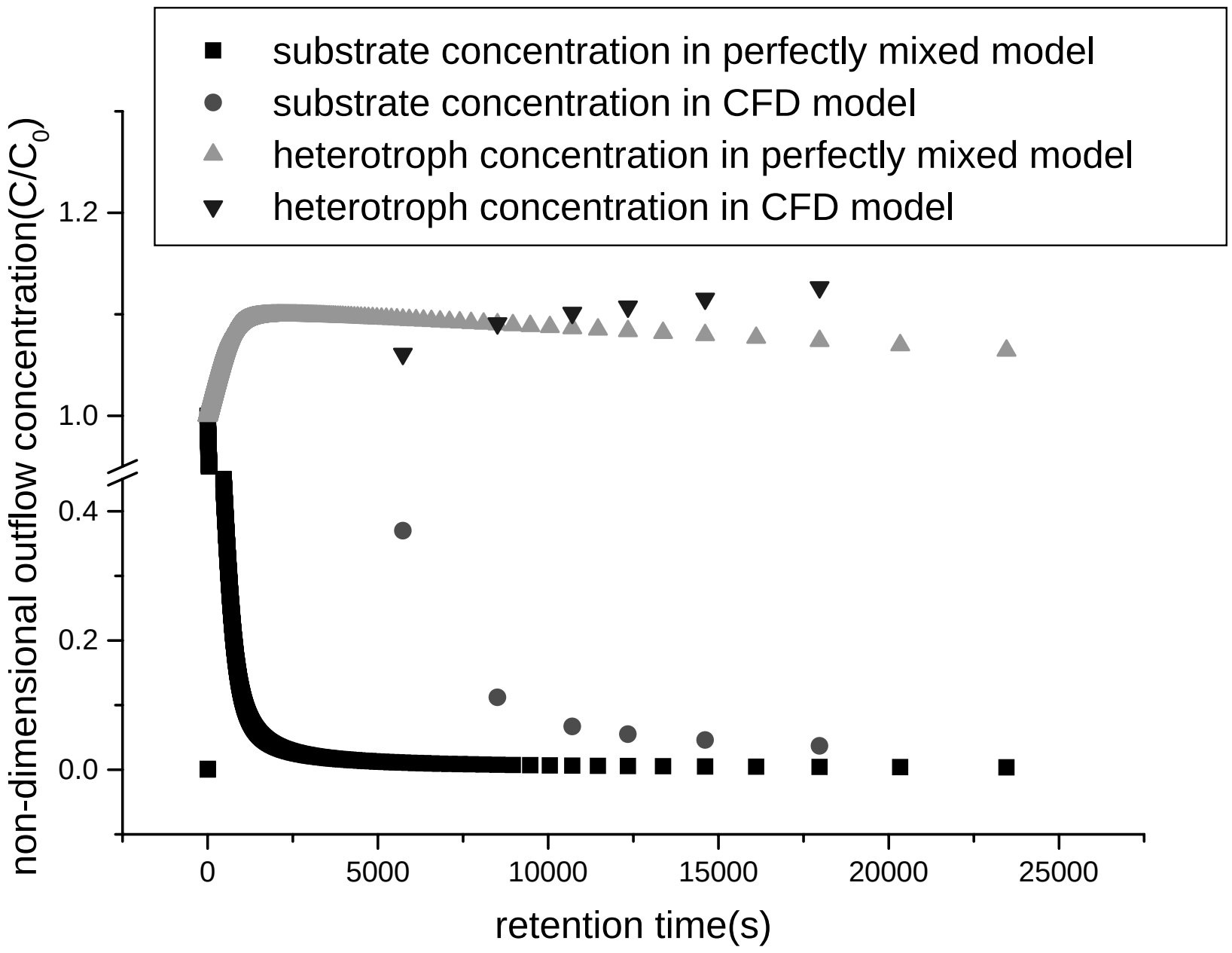

Fig.1 non-dimensional outflow concentration(outflow concentration/inflow concentration) at different retention time simulated by perfectly mixed model and CFD model

Simulation result of the CFD is compared with that of the kinetic model. Figure 1 compares the non-dimensional population of the heterotroph(X/X0) and substrate concentration (C/C0) in outflow in different retention time of the substrate (V/Q) predicted by the perfectly mixed model with that predicted by the CFD model in steady state. It can be seen that the kinetic model predicts a much low substrate concentration in outflow when the same kinetic parameters are used. The kinetic model for biological wastewater treatment system assumes the perfectly mixed state, and often over-predicts the removal efficiency of the substrate in the system[34]. One kind of explanation is the reduction of activity of the heterotrophs in different conditions[35,36]. In comparison with the kinetic model, the CFD model predicts lower performance of the system, because mass transfer effects are considered in CFD approach, that is the main difference of the two kinds of model, so the mass transfer process is the main reason of over-prediction by kinetic model. In the modeling of piston flow reactor for wastewater treatment, dispersion correction[32-33] has also been applied.

The rhological property of biological wastewater treatment system can be regarded as the Newtonian fluid with obviously increased viscosity related to the concentration of heterotrophs[38-39] in CFD calculation. Fig.2 shows the substrate removal efficiency under different viscosity. CFD model predicts that viscosity has low influence on the concentration and substrate in the system as in figure 2.

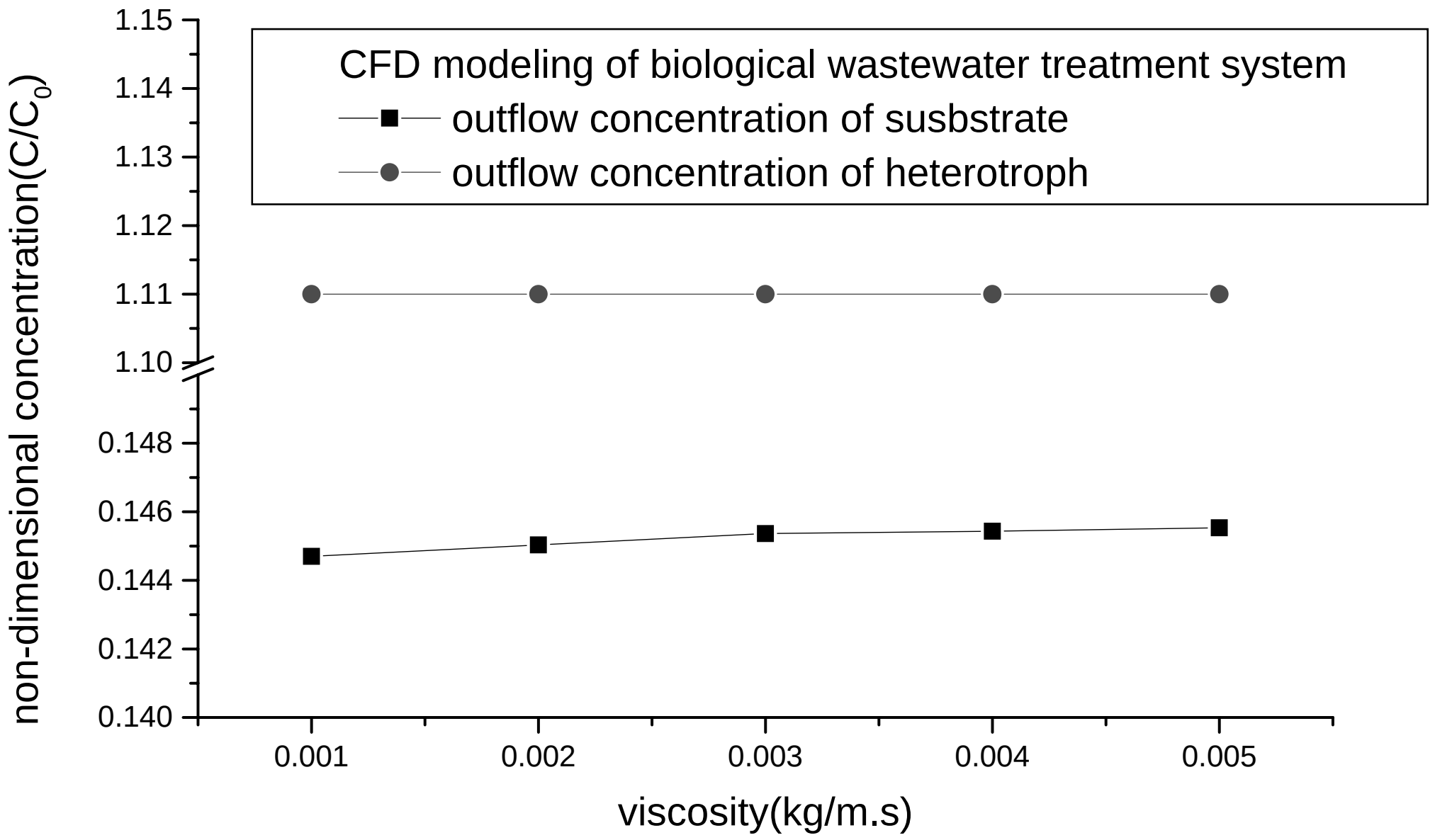

Fig2 non-dimensional concentration of outflow predicted by CFD model with different viscosity

As the kinetic parameters are temperature dependent[23], we compare the performance of the system at different operation temperature. Fig.3 shows the substrate concentration of outflow predicted by CFD model at different operation temperature. It can be seen that the performance of the system decreases at low operation temperature, which is consistent with that of the actual system[23].

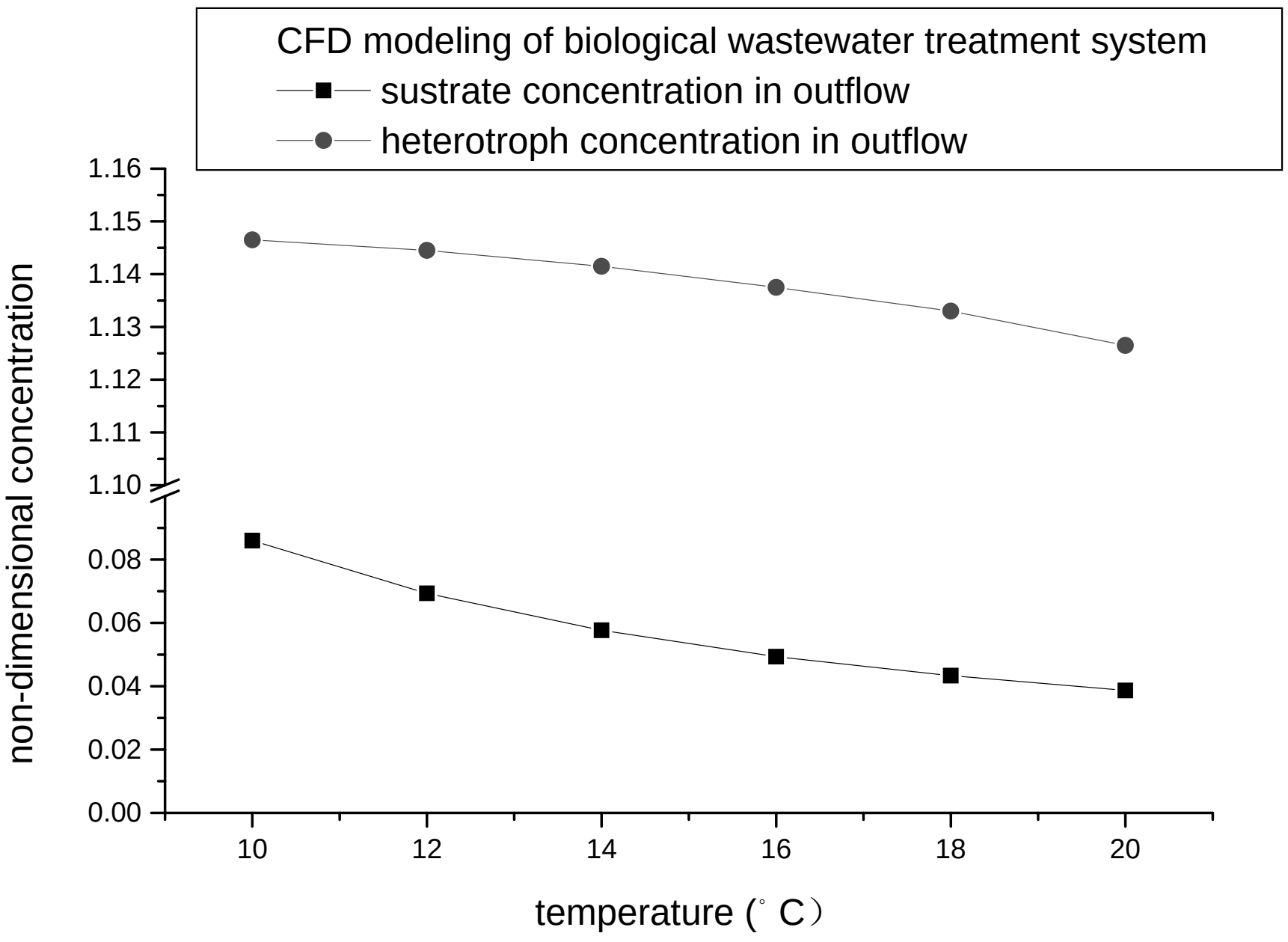

Fig3 non-dimensional concentration of outflow predicted by CFD model under different temperature

Oxygen supply is related to the performance of the system. Figure 4 shows the concentration of the outflow under different oxygen supply. It can be seen that the high oxygen supply greatly improves the performance of the system when at low hydraulic retention time.

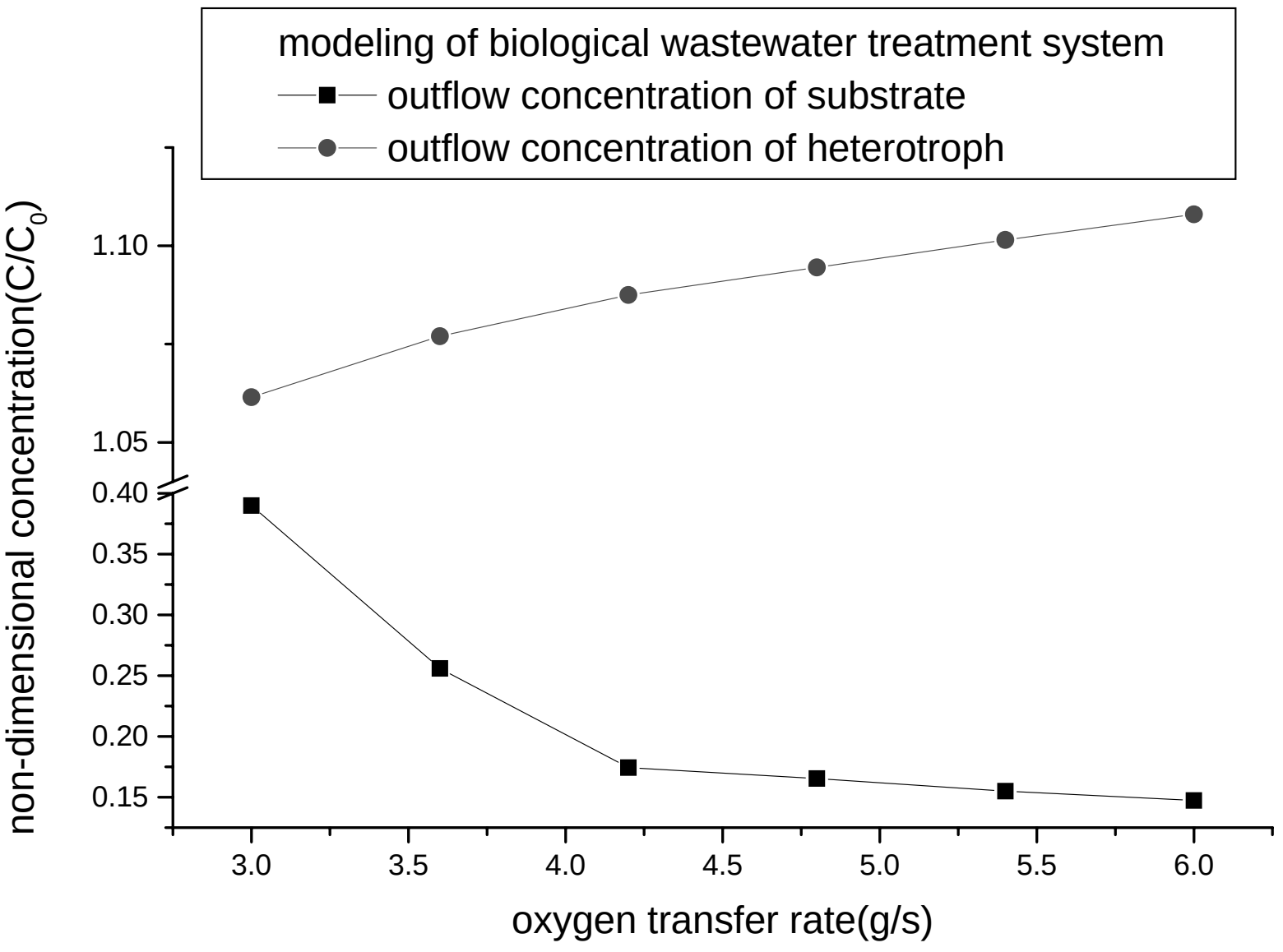

Fig.4 outflow concentration predicted by CFD model under different oxygen supplying rate with 5700s hydraulic retention time

In some of the system, the concentration of the substrate on the surface of the heterotroph is significantly different from the background concentration in the liquid phase. Owing to the size of the heterotroph, there exists a boundary layer for the concentration of substrate on the surface of the heterotroph. When we consider the substrate decomposition reaction rate with equation 2, we should use the substrate concentration on the surface of the heterotroph, however, the transport solution can only give the substrate solution in the liquid. To determine the substrate concentration on the surface of heterotroph, we simplify the heterotroph as a sphere. The mass transfer coefficient k to the surface of a sphere can be found in ref. [40], which can be simplified as

$$kd/D=2. \qquad 12$$

where d is the diameter of sphere, D is the substrate molecular diffusion coefficient in liquid, and thus we have the liquid phase volumetric mass transfer coefficient

$$k_La=kS/V=6k/d=12D/d^2 \qquad 13$$

The transfer rate of the substrate from the liquid phase to the surface of the heterotrophs must be balanced by the removal rate of the substrate in steady state, that is

$$R_d=k_La(C_s-C_{ss})=R(S) \qquad 14$$

where $C_{ss}$ is the concentration of the substrates on the surface of the heterotrophs. $C_s$ is the substrates concentration in liquid phase. $C_{ss}$ can be deduced as:

$$C_{ss}=C_s\frac{\mu_H^0 X_{B,H}/k_La/C_s+K_m/C_s-1}{2}\left(\sqrt{1+\frac{4K_m/C_s}{(\mu_H^0 X_{B,H}/k_La/C_s+K_m/C_s-1)^2}}-1\right) \qquad 15$$

We apply Css to substitute Cs in equation 2 for the modeling to an aerobic granule sludge system[13]. The diameter of the heterotrophs is 4 mm and the substrate diffusion coefficient D is $5*10^{-10}m^2/s$. As a function of heterotrophic maximum specific growth rate, the concentration of substrate can be modeled. The simulated outlet concentrations of substrate and heterotroph are shown in Fig.5. It is found that the performance of the system is not greatly affected by heterotrophic maximum specific growth rate.

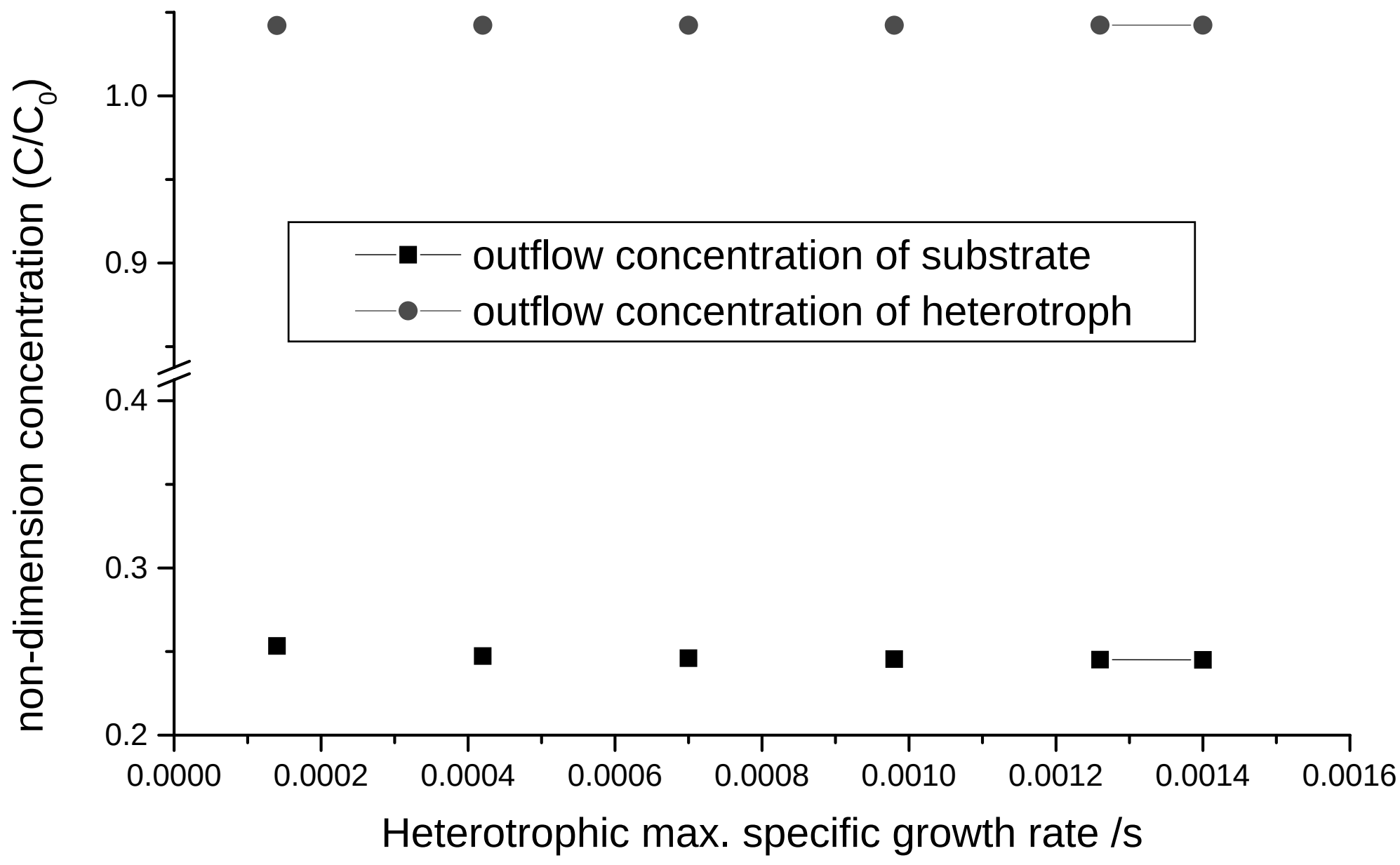

**Figure 5** non-dimensional concentration of outflow predicted by CFD model under different kinetic parameter

Wastewater such as municipal wastewater is very complexes including hundreds and thousands different components. The Activated Sludge Model No. 1(ASM1)[26], developed by the International Water Association (IWA), is considered as the reference kinetic model[14,17]. In ASM1 model, the substrate and microbial are simplified to be 13 components, and 8 biological reactions have been modeled.

The ASM1-based COST benchmark[12] wastewater treatment plant has been specifically developed for simulation-based objective evaluation of different control strategies in wastewater treatment plants. It is based on the ASM1 model, and detailed kinetic formula can be found in COST benchmark homepage[12]. Here we report a prediction of CFD model with inclusion of all the reaction kinetics included in the ASM1 kinetic model. The inlet flow of the CFD model is the same as that of the first tank of the closeloop in COST benchmark, the outflow of the fifth aeration tank predicted by the CFD model is compared with that of the fifth aeration tank in COST benchmark[18]. There are no reaction concerning soluble inert organic matter(SI) and particulate inert organic matter(XI), so their concentration does not vary in steady state reactor.

Table2 the simulation result of CFD comparing with that of COST benchmark

| Outflow | $S_S$ | Xp | $X_S$ | $X_{B,H}$ | $X_{B,A}$ | $S_{NO}$ | $S_{NH}$ | $S_{ND}$ | $X_{ND}$ |
|---|---|---|---|---|---|---|---|---|---|
| COST | 0.808 | 452.74 | 44.48 | 2562.9 | 154.2 | 13.5 | 0.672 | 0.665 | 3.26 |
| CFD | 0.771 | 452.70 | 43.84 | 2563.3 | 154.0 | 12.8 | 0.758 | 0.676 | 3.40 |

Table 3 the parameter difference of CFD model and COST benchmark

| Parameter | $\mu_H$ | $b_H$ | $k_h$ | $\mu_A$ | $b_A$ | $k_a$ | $\eta_g$ |
|---|---|---|---|---|---|---|---|
| COST | 4 | 0.3 | 3 | 0.5 | 0.05 | 0.05 | 0.8 |
| CFD | 49.9 | 0.32 | 3.33 | 24.0 | 0.051 | 0.054 | 0.88 |

The symbols in table 2 are the same as those of COST benchmark[12]. The average dissolved oxygen in reactor is set to be the same as that of COST benchmark through changing $k_La$ of oxygen transfer for each aeration tank. It can be seen from table 2 that CFD model simulates well the biological wastewater treatment system with parameters different from kinetic model (as listed in table3).

**Discussion**

For multi-order reaction, if only one reactant varies remarkably, such as the hydrolysis of particle substrate in biological wastewater treatment system, its behavior is similar to first-order reaction. The first-order reaction rate is not related to the mass transfer of reactants. The main difference between the CFD and kinetic model is the maximum growth rate of microbial as show in table 3. For kinetic model, the maximum growth rate of microbial is very different from the original one. For example, E. coli is one of the main microbial in biological wastewater treatment system, the maximum growth rate of E. coli is 68.6 $day^{-1}$(the generation time is 21minutes). In CFD model of COST benchmark biological wastewater treatment system, for the parameter represents actually the average rate of many different microbial in the system, it is close to the maximum growth rate of E. coli. This result shows the remarkable influence of the mass transfer process to the enzymatic reaction rate, for the difference between the two kinds of parameters is caused by mass transfer process. As many factors, such as the boundary condition around the biological system, influence remarkably the mass transfer of components, and the kinetic model cannot model the influence of these factors, the result implicates that CFD model is more precise than kinetic model in modeling biological system.

Because many enzymes, which come from different biological species and different cells, have the similar structure and catalytic activity, and CFD model use the original kinetic parameter, so the work to study the kinetic information of biological system is remarkably reduced for CFD approach.

In the aerobic granule sludge system, we also show that the removal rate of substrates is not affected by heterotrophic maximum specific growth rate. For a typical particle surface reaction, such as Monad type reaction, Css, the concentration of the substrates on the surface of the particle, is usually far less than $K_m$. According to the Monad reaction kinetics, the removal rate of the substrates is:

$$R(S)= \mu_H^0 X_{B,H} C_{ss}/(K_s+C_{ss}) \approx \mu_H^0 X_{B,H} C_{ss}/ K_s, \qquad 16$$

It is a typical second-order reaction. In steady state, the removal rate R(S) must be compensate by the transport rate $R_d$, which is shown in eq. (14), so

$$C_{ss}/C_s= k_L a /( \mu_H^0 X_{B,H}/ K_s + k_L a) \qquad 17$$

If $k_L a >> \mu_H^0 X_{B,H}/ K_s$, $C_{ss}/C_s \approx 1$, then $R_d=R(S)= \mu_H^0 X_{B,H} C_{ss}/K_s$. In this case, the overall rate of the reaction is controlled by the biological reaction. The opposite extreme condition is $k_L a << \mu_H^0 X_{B,H}/K_s$, $C_{ss}/C_s <<1$, then $R(S)=R_d=k_L a C_s$. In this case, the overall rate of the reaction is controlled by the transport of the reactants. It is also right for particle inner reaction. The above discussed granule sludge system is an example of this condition. For a multi-order particle surface reaction, similar result can be achieved.

As the enzymatic reaction is also a monad-type reaction, we check if an enzymatic reaction in a cell is controlled by the transport of the reactants. It is considered that the substrates transfer to a cell 2r in diameter. The coefficient of substrate mass transfer to the surface of the cell is $k_L a=3D/r^2$. In a typical system with diffusion coefficient D in the order of $10^{-10} m^2/s$, r in the order of 1 um, and typical number of enzymes (N) in a cell is about 1000-10000. Note also that $k_2$ is about $10^3$-$10^6$ /s and $K_m$ is about $10^{-3}$-$10^{-5}$mol/l for typical enzymatic reaction. One can find that the reactant mass transfer to the cell is a much slower process than the kinetic reactions in the cell for most enzymatic reaction. Thus, the reaction rate is often controlled by the transport process in cell. As the biological process in cell is the basic one, such transport controlled enzymatic reaction may be the general phenomena in biological systems.

For many biological reactions, the kinetic parameters are usually difficult to obtain[52]. We have shown that the reaction rate is not greatly influenced by the kinetic parameters in the case of fast reaction. So one of the greatest advantages of the CFD modeling are that remarkable accuracy can be achieved even if we use very rough kinetic parameters.

In comparison with present kinetic models, the CFD model, which is based on the mass transfer process with the inclusion of biological reactions, is a more accurate method for biological system simulation, because more factors influenced the performance of the system are considered in CFD model than in kinetic model. It has been demonstrated in this paper that CFD model can well predict the performance of biological system for wastewater treatment. It has also been shown how CFD methodology can be applied to the study of the complex biological system.

Biological wastewater treatment systems, as well as other biological systems, consist of thousands of simple components and biological reactions. It is not ease to simulate such complicated systems even with chemical kinetic model. However, the system can be simplified to consider the main components and reactions for a specific topic. In modeling substrates and nitrogen removal performance of biological wastewater treatment systems, the difference of many different kinds of substrates and heterotrophs can be ignored, therefore greatly reduces the number of the components and the number of the biochemical reactions. The main factors, which influence the carbon substrate and nitrogen removal efficiency of the system, have been well studied in this paper. Simplifying the chemical reaction brings about errors lesser, and it may be one of the main approaches to save the computer resource for CFD modeling of biological systems. In this paper, less than an hour CPU time on a personal computer equipped with Atholon XP2500 and 1 GB RAM is spended for a complete simulation to the benchmark reactor. In most biological system modeling, less computational resource is needed, for the flow is laminar, and for the calculation of reaction rate is not necessary for many biological systems. With such simplification, reasonable accuracy can be achieved yet the computation time can still be drastically reduced. This simplification can also be used in many biological systems for many other topics.

In bioengineering and bioreactor modeling including biological wastewater treatment reactor modeling, two groups of models have been described in literature[37]. The first type deals with a combination of ideal reactors based on the biological reaction kinetics and the multiple compartments[41-43].The second is based on a three-dimensional network of zones and biological reaction kinetics[44,15-16]. Recently, a combined hybrid multizonal computational fluid dynamics modeling approach[20] has been demonstrated with the CFD tools to obtain mass exchanges between different compartments before dynamical simulation. This kind of combined hybrid multizonal computational fluid dynamics modeling approach obtains substantial improvement on solving the difficulties of characterizing the mass and energy fluxes between adjacent zones for the multizone model. However, the combined multizone computational fluid dynamics modeling approach still uses the completely mixed hypothesis in each zone, and would inevitably make some errors in industrial reactor simulation, although the errors can be greatly reduced through increasing the number of the zones, but then the model construction becomes rather sophisticated. In this paper, we have successfully demonstrated that a CFD model with the inclusion of the biological reaction kinetics can be used. It gives a more accurate description of the processes in the bioreactor, and can be applied to modeling bioreactor with other biological processes. The CFD approach can be also applied to modeling water ecologic system such as marine ecosystem and water eutrophication simulation.

## Acknowledgement

This work is partially supported by the Ministry of Science and Technology of the People's Republic of China (grants No. 2003AA60100-3-4). The authors are grateful to Mr. Xiaoyu Yang in Shanghai Hi-key technology corporation Ltd for his technical support and advices to use Fluent.